\documentclass[11pt]{article}

\usepackage{times}
\usepackage{xspace}
\usepackage{amsmath}
\usepackage{amssymb}
\usepackage{amsfonts}
\usepackage{latexsym}
\usepackage{todonotes}
\usepackage{booktabs}
\usepackage{enumerate}
\usepackage[nodate]{datetime}
\usepackage{verbatim}

\newtheorem{xdefinition}{Definition}
\newtheorem{xobservation}{Observation}
\newtheorem{xtheorem}{Theorem}
\newtheorem{xlemma}{Lemma}
\newtheorem{xproposition}{Proposition}
\newtheorem{xcorollary}{Corollary}
\newenvironment{definition}{\begin{xdefinition}\rm}%
{\hspace*{\fill}\raisebox{-1pt}{\boldmath$\Box$}\end{xdefinition}}
{\hspace*{\fill}\raisebox{-1pt}{\boldmath$\Box$}\end{xobservation}}

{\hspace*{\fill}\raisebox{-1pt}{\boldmath$\Box$}\end{trivlist}}

\newcommand{\OPT}{\ensuremath{\operatorname{\textsc{Opt}}}\xspace}
\newcommand{\opt}{\OPT}
\newcommand{\algA}{{\ensuremath{\mathbb{A}}}\xspace}
\newcommand{\algB}{{\ensuremath{\mathbb{B}}}\xspace}
\newcommand{\algC}{{\ensuremath{\mathbb{C}}}\xspace}
\newcommand{\algM}{{\ensuremath{\mathbb{M}}}\xspace}
\newcommand{\ALG}{{\ensuremath{\mathbb{A}}}\xspace}
\newcommand{\alg}{{\ensuremath{\mathbb{A}}}\xspace}

\newcommand{\SET}[1]{\left\{#1\right\}}
\newcommand{\SETOF}[2]{\SET{#1 \mid #2}}

\newcommand{\SEQ}[1]{\left\langle #1\right\rangle}

\newcommand{\firstfit}{\ensuremath{\operatorname{\textsc{First-Fit}}}\xspace}
\newcommand{\bestfit}{\ensuremath{\operatorname{\textsc{Best-Fit}}}\xspace}

\newcommand{\worstfit}{\ensuremath{\operatorname{\textsc{Worst-Fit}}}\xspace}
\newcommand{\FFD}{\ensuremath{\operatorname{\textsc{First-Fit-Decreasing}}}\xspace}
\newcommand{\FFI}{\ensuremath{\operatorname{\textsc{First-Fit-Increasing}}}\xspace}
\newcommand{\LRU}{\ensuremath{\operatorname{\textsc{LRU}}}\xspace}
\newcommand{\LRUtwo}{\ensuremath{\operatorname{\textsc{LRU-2}}}\xspace}
\newcommand{\FIFO}{\ensuremath{\operatorname{\textsc{FIFO}}}\xspace}

\newcommand{\FWF}{\ensuremath{\operatorname{\textsc{FWF}}}\xspace}
\newcommand{\RLRU}{\ensuremath{\operatorname{\textsc{RLRU}}}\xspace}

\newcommand{\movetofront}{\ensuremath{\operatorname{\textsc{Move-To-Front}}}\xspace}
\newcommand{\frequencycount}{\ensuremath{\operatorname{\textsc{Frequency-Count}}}\xspace}
\newcommand{\transpose}{\ensuremath{\operatorname{\textsc{Transpose}}}\xspace}
\newcommand{\timestamp}{\ensuremath{\operatorname{\textsc{Timestamp}}}\xspace}
\newcommand{\MTF}{\ensuremath{\operatorname{\textsc{MTF}}}\xspace}
\newcommand{\FC}{\ensuremath{\operatorname{\textsc{FC}}}\xspace}
\newcommand{\T}{\ensuremath{\operatorname{\textsc{Trans}}}\xspace}
\newcommand{\greedyfit}{\ensuremath{\operatorname{\textsc{Greedy-Fit}}}\xspace}
\newcommand{\onebin}{\ensuremath{\operatorname{\textsc{One-Bin}}}\xspace}
\newcommand{\postgreedy}{\ensuremath{\operatorname{\textsc{Post-Greedy}}}\xspace}
\newcommand{\fast}{\ensuremath{\operatorname{\textsc{Fast}}}\xspace}
\newcommand{\WF}{\ensuremath{\operatorname{\textsc{WF}}}\xspace}
\newcommand{\NF}{\ensuremath{\operatorname{\textsc{NF}}}\xspace}
\newcommand{\DC}{\ensuremath{\operatorname{\textsc{DC}}}\xspace}
\newcommand{\LDC}{\ensuremath{\operatorname{\textsc{LDC}}}\xspace}
\newcommand{\GREEDY}{\ensuremath{\operatorname{\textsc{Greedy}}}\xspace}

\newcommand{\cratio}[1]{\ensuremath{\mathrm{CR}_{#1}}\xspace}
\newcommand{\mmratio}[1]{\ensuremath{\mathrm{MR}_{#1}}\xspace}
\newcommand{\roratio}[1]{\ensuremath{\mathrm{RR}_{#1}}\xspace}

\newcommand{\rwor}[2]{\ensuremath{\mathrm{WR}_{#1,#2}}\xspace}

\begin{document}

\title{Relative Worst-Order Analysis: A Survey\thanks{Supported in part by the 
Independent Research Fund Denmark, Natural Sciences,
grants DFF-7014-00041.}
}
\author{Joan Boyar \hspace*{2em}
        Lene M. Favrholdt \hspace*{2em}
        Kim S. Larsen 
\\[1ex]
{\normalsize University of Southern Denmark, Odense, Denmark} \\
{\normalsize \texttt{\{joan,lenem,kslarsen\}@imada.sdu.dk}}
}

\date{}

\maketitle

\begin{abstract}
  Relative worst-order analysis is a technique for assessing the
  relative quality of online algorithms. We survey the
  most important results obtained with this technique
  and compare it with other quality measures.
\end{abstract}

\section{Introduction}
Online problems are optimization problems where the input arrives one request at
a time, and each request must be processed without knowledge of
future requests.
The investigation of online algorithms was largely initiated by the
introduction of competitive analysis by Sleator and Tarjan~\cite{ST85}.
They introduced the method as a general analysis technique,
inspired by approximation algorithms. The term ``competitive'' is
from Karlin et al.~\cite{KMRS88} who named the worst-case ratio of
the performance of the online to the offline algorithm the
``competitive ratio''.
Many years earlier, Graham carried out what is now viewed as an
example of a competitive analysis~\cite{G69}.

The over-all goal of a theoretical quality measure is to predict behavior
of algorithms in practice.
In that respect, competitive analysis works well in some cases,
but, as pointed out by the inventors~\cite{ST85} and others, fails
to discriminate between good and bad algorithms in other cases.
Ever since its introduction, researchers have worked on improving the measure,
defining variants, or defining measures based on other concepts to improve
on the situation.
Relative worst-order analysis (RWOA), a technique for assessing the
relative quality of online algorithms, is
one of the most thoroughly tested such proposals. 

RWOA was originally defined by Boyar
and Favr\-holdt~\cite{BF07}, and the definitions were extended
together with Larsen~\cite{BFL07}.
As for all quality measures, an important issue is to be able
to separate algorithms, i.e., determine which of two
algorithms is the best.
RWOA has been shown to be applicable
to a wide variety of problems and provide separations,
not obtainable using competitive analysis,
corresponding better to experimental results or intuition
in many cases.

In this survey, we motivate and define RWOA, outline the background for its
introduction, survey the most important results, and compare it to
other measures.

\section{Relative Worst-Order Analysis}
As a motivation for RWOA,
consider the following desirable property of a quality measure for
online algorithms:
For a given problem $P$ and two algorithms \algA and \algB for
$P$, if \algA performs at least as well as \algB on every possible
request sequence and better on many, then the quality measure indicates
that \algA is better than \algB.
We consider an example of such a situation for the paging problem.

\subsection{A Motivating Example}
\label{sec:motivating_ex}
In the paging problem, there is a cache with $k$ pages and a larger,
slow memory with $N>k$ pages. The request sequence consists
of page numbers in $\SET{1,\ldots ,N}$.
When a page is requested, if it is not
among the at most $k$ pages in cache, there is a fault, and
the missing page must be brought into cache. If the cache is full, this means
that some page must be evicted from the cache. The goal is to minimize
the number of faults. Clearly, the only thing we can control algorithmically
is the eviction strategy.

We consider two paging algorithms,
\LRU (Least-Recently-Used) and \FWF (Flush-When-Full).
On a fault with a full cache, \LRU always evicts its
least recently used page from cache. \FWF, on the other hand, evicts
everything from cache in this situation. It is easy to see
that, if run on the same sequence, whenever \LRU faults, \FWF also faults,
so \LRU performs at least as well as \FWF on every sequence. \LRU usually faults
less than \FWF~\cite{Y94}.
It is well known that \LRU and
\FWF both have competitive ratio~$k$, so competitive analysis does not
distinguish between them, and there are relatively few measures which do.
RWOA, however, is one such measure~\cite{BFL07}.
In Section~\ref{LRUvsFWF}, we consider \LRU and \FWF in greater detail to
give a concrete example of RWOA.

\subsection{Background and Informal Description}
Table~\ref{measures} gives informal ``definitions'' of 
the relative worst-order ratio and related measures.
The ratios shown in the table capture the general ideas,
although they do not reflect that the measures are asymptotic measures.
We discuss the measures below, ending with a formal definition of the
relative worst-order ratio.

\begin{table}[t]
\begin{center}
\begin{tabular}{l@{\hspace{2em}}l}
\toprule
Measure & Value \\
\midrule
\raisebox{0ex}[4.5ex][3.5ex]{\mbox{}} 
Competitive ratio &
\: $\displaystyle \cratio{\algA} = \sup\limits_{I} \frac{\algA(I)}{\opt(I)}$ \\
\raisebox{0ex}[6ex][3.5ex]{\mbox{}} 
Max/max ratio & \: $\displaystyle \mmratio{\algA} =  \frac{\max\limits_{|I|=n}             
\algA(I)}{\max\limits_{|I'|=n} \opt({I'})}$ \\
\raisebox{0ex}[6ex][3.5ex]{\mbox{}} 
Random-order ratio & \: $\displaystyle\roratio{\algA} = \sup\limits_{I}                   
\frac{E_{\pi}\big[\algA(\pi(I))\big]}{\opt(I)}$ \\
\raisebox{0ex}[6ex][3.5ex]{\mbox{}} 
Relative worst-order ratio \; & \: $\displaystyle\rwor{\algA}{\algB} = \sup\limits_{I}    
\frac{\sup\limits_{\pi}\big\{\algA(\pi(I))\big\}}
{\sup\limits_{\pi'}\big\{\algB(\pi'(I))\big\}}$ \;
\\
\bottomrule
\end{tabular}
\end{center}
\caption{\label{measures}
Simplified ``definitions'' of  measures}
\end{table}

RWOA compares two online algorithms directly,
rather than indirectly by first comparing both
to an optimal offline algorithm.
When differentiating between online algorithms is the goal, performing a
direct comparison between the algorithms can be an advantage; first
comparing both to an optimal offline algorithm and then comparing the
results, as many performance measures including competitive analysis do,
can lead to a loss of information. This appears to be
at least part of the problem when comparing \LRU to \FWF with
competitive analysis, which finds them equally bad.
Measures comparing directly,
such as RWOA, bijective and average analysis~\cite{ADLO07},
and relative interval analysis~\cite{DLM09}, would generally indicate correctly
that \LRU is the better algorithm.

Up to permutations of the request sequences,
if an algorithm is always at
least as good and sometimes better than another, RWOA separates them.
RWOA compares two algorithms on their respective worst
orderings of sequences having the same content.
This is different from competitive analysis where an
algorithm and \opt are compared on the same sequence. 
When comparing
algorithms directly, using exactly the same sequences will tend to produce
the result that many algorithms are not comparable, because one
algorithm does well on one type of sequence, while the other
does well on another type. In addition, comparing on possibly different
sequences can make it harder for the adversary to produce unwanted,
pathological sequences which may occur seldom in practice, but skew
the theoretical results.
Instead, with RWOA, online algorithms are compared directly to each 
other on their respective worst permutations of the request sequences. 
This comparison in RWOA combines some of the desirable properties of the
max/max ratio~\cite{BB94} and the random-order ratio~\cite{K96}.

\subsubsection{The Max/Max Ratio}
With the max/max ratio defined by Ben-David and Borodin,
 an algorithm is compared
to \opt on its and \opt's respective worst-case sequences of the same length.
Since \opt's worst sequence of any given length is the same, regardless
of which algorithm it is being compared to,
comparing two online algorithms directly gives the same
result as dividing their max/max ratios.
Thus, the max/max ratio allows direct comparison of two online algorithms,
to some extent, without the intermediate comparison to \opt.
The max/max ratio can only provide interesting results
when the length of an input sequence yields a
bound on the cost/profit of an optimal solution.

In the paper~\cite{BB94} introducing the max/max ratio, the $k$-server
problem is analyzed.
This is the problem where $k$ servers are placed in a metric space,
and the input is a sequence of requests to points in that space.
At each request, a server must be moved to the requested point
if there is not already a server at the point.
The objective is to minimize the total distance the servers are moved.
It is demonstrated that, for $k$-server on a bounded metric space, the
max/max ratio can provide more optimistic and detailed results than competitive analysis.
Unfortunately, there is still the loss of information as generally
occurs with the indirect comparison to \opt,
and the max/max ratio does not distinguish between \LRU and \FWF, or actually
between any two deterministic online paging algorithms. 

However, the possibility of directly comparing online
algorithms and comparing them on their respective worst-case sequences
from some partition of the space of request sequences was inspirational.
RWOA uses a more fine-grained partition than partitioning with respect to
the sequence length.
The idea for the specific partition used stems from the random-order ratio.

\subsubsection{The Random-Order Ratio}
The random-order ratio was introduced in~\cite{K96} by Kenyon (now Mathieu).
The appeal of this quality measure is that it allows considering some
randomness of the input sequences without specifying a complete
probability distribution.
It was introduced in connection with bin
packing, i.e., the problem of packing items of sizes between
$0$ and~$1$ into as few bins of size~$1$ as possible.
For an algorithm \algA for this minimization problem,
the random-order
ratio is the maximum ratio, over all multi-sets of items,
of the expected
performance, over all permutations of the multi-set, of \algA compared with an optimal solution; see also Table~\ref{measures}.
If, for all possible multi-sets of items, any permutation of these items is
equally likely, this ratio gives a meaningful worst-case measure of how well
an algorithm can perform.

In the paper introducing the random-order ratio,
it was shown that for bin packing, 
the random-order ratio of \bestfit lies between $1.08$ and $1.5$.
In contrast, the competitive ratio of \bestfit is $1.7$~\cite{JDUGG74}.

Random-order analysis has also been applied to other problems, e.g.,
knapsack~\cite{BIKK07},
bipartite matching~\cite{GM08,DH09}, scheduling~\cite{OT08,GKT15},
bin covering~\cite{CFL14,FR16}, and facility location~\cite{M01}.
However, the analysis is often rather challenging, and
in~\cite{CCRZ08}, a simplified version of the random-order ratio is
used for bin packing.

\subsection{Definitions}
Let $I$ be a request sequence of length $n$ for an online problem
$P$. 
If $\pi$ is a permutation on $n$ elements, then $\pi(I)$ denotes
$I$ permuted by $\pi$.

If $P$ is a minimization problem, $\alg(I)$ denotes the cost of the
algorithm \alg on the sequence $I$, and $$\alg_W(I)=\max_{\pi}
\alg(\pi(I)),$$
where $\pi$ ranges over the set of all permutations of $n$ elements.

If $P$ is a maximization problem, $\alg(I)$ denotes the profit of the
algorithm \alg on the sequence $I$, and $$\alg_W(I)=\min_{\pi}
\alg(\pi(I)).$$

Informally, RWOA compares two algorithms,
\algA and \algB, by partitioning the set of
request sequences as follows: Sequences are in the same part of the
partition if and only if they are permutations of each other.
The relative worst-order ratio is defined for algorithms
\algA and \algB, whenever one algorithm performs at least as well
as the other on every part of the partition, i.e., whenever
$\algA_W(I)\leq \algB_W(I)$, for all request sequences~$I$, or
$\algA_W(I)\geq \algB_W(I)$, for all request sequences~$I$ (in the
definition below, this corresponds to $c_u(\algA,\algB) \leq 1$ or $c_l(\algA,\algB) \geq 1$). 
In this case,
to compute the relative worst-order ratio of \algA to \algB,
we compute a bound ($c_l(\algA,\algB)$ or $c_u(\algA,\algB)$) on the ratio of how the two 
algorithms perform on their respective worst permutations of some sequence.
Note that the two algorithms may have different worst permutations
for the same sequence. 

We now state the formal definition:
\begin{definition}\label{def:rwor}
For any pair of algorithms $\algA$ and $\algB$, we define
\[\begin{array}{rcrl}
  c_l(\algA,\algB) &=&
    \sup & \SETOF{c}{\exists b\, \forall I \colon \algA_W (I) \geq c \, \algB_W(I) - b} \nonumber
  \mbox{ and } \\
  c_u(\algA,\algB) &=&
    \inf & \SETOF{c}{\exists b\, \forall I \colon \algA_W (I) \leq c \, \algB_W(I) + b}. \nonumber
\end{array}\]
If $c_l(\algA,\algB) \geq 1$ or $c_u(\algA,\algB) \leq 1$, the 
algorithms are said to be {\em comparable}
and the {\em relative worst-order ratio}
$\rwor{\algA}{\algB}$ of algorithm $\algA$ to algorithm $\algB$ is
defined as
$$
\rwor{\algA}{\algB} = 
\begin{cases}
 c_u(\algA,\algB), & \text{ if } c_l(\algA,\algB) \geq 1, \text{ and}\\
 c_l(\algA,\algB), & \text{ if } c_u(\algA,\algB) \leq 1.
\end{cases}
$$
Otherwise, $\rwor{\algA}{\algB}$ is undefined.

For a minimization (maximization) problem, the algorithms $\algA$ and
$\algB$ are said to be {\em comparable in $\algA$'s favor} if
$\rwor{\algA}{\algB} < 1$ ($\rwor{\algA}{\algB} > 1$).
Similarly, the algorithms are
said to be {\em comparable in $\algB$'s favor},
if $\rwor{\algA}{\algB} > 1$ ($\rwor{\algA}{\algB} < 1)$.
\end{definition}

Note that the ratio $\rwor{\algA}{\algB}$ can be larger than or smaller than one depending
on whether the problem is a minimization problem or a maximization problem
and which of \algA and \algB is the better algorithm.
Table~\ref{interpretation} indicates the result in each case.
\begin{table}[hbt]
\begin{center}
\begin{tabular}{l@{\hspace{2em}}c@{\hspace{2em}}c}
\toprule
Result & Minimization & Maximization \\
\midrule
$\algA$ better than $\algB$ & $<1$ & $>1$ \\
$\algB$ better than $\algA$ & $>1$ & $<1$ \\
\bottomrule
\end{tabular}
\end{center}
\caption{\label{interpretation}
  Relative worst-order ratio interpretation, depending on whether the
problem is a minimization or a maximization problem.}
\end{table}
Instead of saying that two algorithms, \algA and \algB, are comparable
in \algA's favor, one would often just say that \algA is better than
\algB according to RWOA.

For quality measures evaluating algorithms by comparing them to each
other directly, it is particularly important to be transitive: 
If \algA and \algB are comparable
in \algA's favor and \algB and \algC are comparable in \algB's
favor, then \algA and \algC are comparable in \algA's favor.
When this transitivity holds, to prove that a new algorithm is
better than all previously known algorithms, one only has to
prove that it is better than the best among them. This holds for
RWOA~\cite{BF07}.

\section{Paging}

In this section, we survey the most important RWOA results for
paging and explain how they differ from the results obtained
with competitive analysis. 
As a relatively simple, concrete example of RWOA, we first explain how
to obtain the separation of \LRU and \FWF~\cite{BFL07} mentioned in
Section~\ref{sec:motivating_ex}.  

\subsection{\LRU vs.\ \FWF}
\label{LRUvsFWF}
The first step in computing
the relative worst-order ratio, $\rwor{\FWF}{\LRU}$, is to show that
\LRU and \FWF
are comparable. Consider any request sequence~$I$
for paging with cache size~$k$. For
any request $r$ to a page $p$ in~$I$, if \LRU faults on $r$, either
$p$ has never been requested before or there have been at least
$k$ different requests to distinct pages other than $p$ since the
last request to~$p$. In the case where $p$ has never been requested
before, any online algorithm faults on~$r$. If there have been
at least $k$ requests to distinct pages other than $p$ since the
last request to~$p$, \FWF has flushed since that last request to $p$,
so $p$ is no longer in its cache and \FWF faults, too. Thus, for any
request sequence $I$, $\FWF(I)\geq \LRU(I)$. Consider \LRU's worst
ordering, $I_{\LRU}$, of a sequence $I$. Since \FWF's performance
on its worst ordering of any sequence is at least as bad as its
performance on the sequence itself, $\FWF_W(I_{\LRU})\geq \FWF(I_{\LRU})\geq 
\LRU(I_{\LRU}) = \LRU_W(I_{\LRU})$. Thus, $c_l(\FWF,\LRU)\geq 1$.

As a remark, in general, to prove that one algorithm is at least as good as
another on their respective worst orderings of all sequences, one
usually starts with an arbitrary sequence and its worst ordering
for the better algorithm. Then, that sequence is gradually permuted,
starting at the beginning, so that the poorer algorithm does at least
as badly on the permutation being created.

The second step is to show the separation, giving a lower bound on
the term $c_u(\FWF,\LRU)$.
We assume that the cache is initially empty.
Consider the sequence $I_s = \SEQ{1,2,\ldots,k,k+1,k,\ldots,2}^s$, where
\FWF faults on all $2ks$ requests. 
\LRU only faults on $2s+k-1$ requests in all, the first $k$ requests and
every request to $1$ or $k+1$ after that, but we need to
consider how many times \LRU faults on its worst ordering of $I_s$.

It is proven in~\cite{BFL07} that, for any sequence $I$, there is
a worst ordering of $I$ for \LRU that has all faults before all
hits (requests which are not faults). The idea is to consider any
worst order of $I$ for \LRU and move requests which are hits, but
are followed by a fault towards the end of the sequence without
decreasing the number of faults. Since \LRU needs $k$ distinct
requests between two requests to the same page in order to fault,
with only $k+1$ distinct pages in all, the faults at the beginning
must be a cyclic repetition of the $k+1$ pages. 
Thus, a worst ordering of $I_s$
for \LRU is $I'_s = \SEQ{2,3,\ldots,k,k+1,1}^s,\SEQ{2,\ldots,k}^s$,
and $\LRU(I'_s)=(k+1)s+k-1$. This means that, asymptotically,
$c_u(\FWF,\LRU)\geq \frac{2k}{k+1}$. We now know that $\rwor{\FWF}{\LRU}
\geq \frac{2k}{k+1}$, showing that \FWF and \LRU are comparable in
\LRU's favor, which is the most interesting piece of information.

However, one can prove that this is the exact result.
In the third step, we prove that $c_u(\FWF,\LRU)$ cannot be larger
than $\frac{2k}{k+1}$, asymptotically.
In fact, this is shown in~\cite{BFL07} by proving the more general
result that, for any \emph{marking} algorithm~\cite{BIRS95}, $\algM$, and for any request
sequence $I$, $\algM_W(I) \leq \frac{2k}{k+1}\LRU_W(I)+k$. A marking
algorithm is defined with respect to \emph{$k$-phases}, a partitioning
of the request sequence. Starting at the beginning of $I$, the
first phase ends with the request immediately preceding the $(k+1)$st
distinct page, and succeeding phases are also longest intervals
containing at most $k$ distinct pages.
An algorithm is a marking algorithm if,
assuming we mark a page each time it is requested and
start with no pages marked at the beginning of each phase,
the algorithm never evicts a marked page.
As an example, \FWF is a marking algorithm.
Now, consider any sequence, $I$,
with $m$ $k$-phases. A marking algorithm $\algM$ faults at most $km$ times
on $I$. Any two consecutive $k$-phases in $I$ contain at least $k+1$ pages,
so there must be a permutation of the sequence where \LRU faults
at least $k+1$ times on the requests of each of the $\lfloor \frac{m}{2}
\rfloor$ consecutive pairs of $k$-phases in $I$. This gives the desired asymptotic
upper bound, showing that $\rwor{\FWF}{\LRU}=\frac{2k}{k+1}$.

\subsection{Other Paging Algorithms}
\label{venividivici}

Like \LRU and \FWF, the algorithm \FIFO also has competitive
ratio~$k$~\cite{BE98}.
\FIFO simply evicts the first page that entered the cache, regardless
of its use while in cache.
In experiments, both \LRU and \FIFO are consistently much better
than \FWF. 
\LRU and \FIFO are both {\em conservative} algorithms~\cite{Y94}, meaning that on
any sequence of requests to at most $k$ different pages, each of them
faults at most $k$ times.
This means that, according to RWOA, they are equally good and both are
better than \FWF, since for any
pair of conservative paging algorithms, \algA and \algB,
$\rwor{\algA}{\algB}=1$ and
$\rwor{\FWF}{\algA} = \frac{2k}{k+1}$~\cite{BFL07}.

With a quality measure that separates \FWF and \LRU, an obvious question
to ask is: Is there a paging algorithm which is better than \LRU according
to RWOA?
The answer to this is ``yes''.
\LRUtwo~\cite{OOW93}, which was proposed for database disk buffering,
is the algorithm which evicts the page with the earliest second-to-last request.
\LRUtwo and \LRU are $(1+\frac{1}{2k+2},\frac{k+1}{2})$-related.
This concept was introduced in~\cite{BFL07}, expressing that
$c_u(\LRUtwo,\LRU)=1+\frac{1}{2k+2}$ and 
$c_u(\LRU,\LRUtwo)=\frac{k+1}{2}$
(see Definition~\ref{def:rwor} for a definition of $c_u$).
Thus, the algorithms are \emph{asymptotically comparable} in \LRUtwo's
favor~\cite{BEKL10}.

In addition, a new algorithm, \RLRU (Retrospective \LRU),
was defined in~\cite{BFL07} and
shown to be better than \LRU according to RWOA.
Experiments, simply comparing the number of page faults on the same
input sequences, have shown that \RLRU is consistently
slightly better than \LRU~\cite{MN12}.
\RLRU is a phase-based algorithm. When considering a request, it
determines whether
\OPT would have had the page in cache given the sequence
seen so far (this is efficiently computable), and uses that information
in a marking procedure.

Interestingly, \LRUtwo has competitive ratio $2k$ and \RLRU has competitive
ratio $k+1$, so both are worse than \LRU according to competitive analysis.

Also for paging, considering \LRU and $\LRU(\ell)$, which is \LRU adapted
to use look-ahead $\ell$ (the next $\ell$ requests after the current one), 
evicting a least recently used page \emph{not}
occurring in the look-ahead, both algorithms have competitive ratio~$k$,
though look-ahead helps significantly in practice.
Using RWOA, $\rwor{\LRU}{\LRU(\ell)} = \min\SET{k, \ell+1}$,
so $\LRU(\ell)$ is better~\cite{BFL07}.

\section{Other Online Problems}

In this section, we give further examples of problems and algorithms
where RWOA gives results that are qualitatively different from those
obtained with competitive analysis.
We consider various problems,
including list accessing, bin packing, bin coloring, and scheduling. 

List accessing~\cite{ST85,AW98} is a
classic problem in data structures, focusing on maintaining an optimal
ordering in a linked list.
In online algorithms, it also has the r\^{o}le of a
theoretical benchmark problem, together with paging
and a few other problems,
on which many researchers evaluate new techniques or quality measures.

The problem is defined as follows:
A list of items is given and requests are to items in the list.
Treating a request requires accessing the item, and the cost
of that access is the index of the item, starting with one.
After the access, the item can be moved to any location closer
to the front of the list at no cost.
In addition, any two consecutive items may be transposed at a cost of one.
The objective is to minimize the total cost of processing the
input sequence.

We consider three list accessing algorithms:
On a request to an item $x$, the algorithm \movetofront (\MTF)~\cite{M65}
moves $x$ to the front of the list, whereas the algorithm \transpose
(\T) just swaps $x$ with its predecessor. The third algorithm,
\frequencycount (\FC), keeps the list sorted by the number of times
each item has been requested.

For list accessing~\cite{ST85}, letting $l$ denote the length of the list,
the algorithm \movetofront
has strict competitive ratio $2-\frac{2}{l+1}$~\cite{I91}
(referring to personal communication, Irani credits Karp and Raghavan
 with the lower bound).
In contrast, \frequencycount and \transpose both
have competitive ratio $\Omega(l)$~\cite{BE98}.
Extensive experiments demonstrate that \MTF and \FC are approximately equally
good, whereas \T is much worse~\cite{BM85,BE97}.
Using RWOA, \MTF and \FC are equally good, whereas both
$\rwor{\T}{\MTF}\in\Omega(l)$ and $\rwor{\T}{\FC}\in\Omega(l)$,
so \T is much worse~\cite{EKL13}.

For bin packing, both Worst-Fit (\WF), which places an item
in a bin with largest available space (but never opens a new bin
unless it has to), and Next-Fit (\NF), which closes
its current bin whenever an item does not fit (and never considers
that bin again), have competitive ratio~$2$~\cite{J74}.
However, \WF is at least as good as \NF on every sequence
and sometimes much better~\cite{BEL10}.
Using RWOA, $\rwor{\NF}{\WF}=2$, so \WF is the better algorithm.

Bin coloring is a variant of bin packing, where items are unit-sized
and each have a color. The goal is to minimize the 
maximum number of colors in any bin, under the restriction that only a
certain number, $q$, of bins are allowed to be open at any time and a bin
is not closed until it is full.
Consider the algorithms \onebin, which never has more than one
bin open, and \greedyfit, which always keeps $q$ open
bins, placing an item in a bin already having that color, if possible, and
otherwise in a bin with fewest colors.
We claim that \greedyfit is obviously the better algorithm, but if the
bin size is larger than approximately $q^3$, 
\onebin has a better competitive ratio than \greedyfit~\cite{KPRS08}.
However, according to 
 RWOA,
 \greedyfit is better~\cite{EFK12}.

For Scheduling on two related machines to minimize makespan (the time
when all jobs are completed),
the algorithm \fast, which only uses the fast machine, is
$\frac{s}{s+1}$-competitive, where $s$ is the speed ratio of the two
machines.
If $s$ is larger than the golden ratio, this is the best possible
competitive ratio.
However, the algorithm \postgreedy, which schedules each job on the
machine where it would finish first, is never worse than \fast and
sometimes better. This is reflected in the relative worst-order
ratio, since $\rwor{\fast}{\postgreedy} = \frac{s+1}{s}$~\cite{EFK06}.

In addition to these examples, it is widely believed and consistent
with experiments that for bin packing problems,
\firstfit algorithms perform better than \worstfit algorithms,
and that processing larger items first is better than processing
smaller items first.
For the problem examples below, competitive analysis cannot
distinguish between the algorithms, that is, they have
the same competitive ratio, whereas using RWOA, we
get the separation in the right direction.
The examples are the following:
For dual bin packing (the variant of bin packing where there is a fixed
number of bins, the aim is to pack as many items as possible, and all
bins are considered open from the beginning), \firstfit is better than \worstfit~\cite{BF07}.
For grid scheduling (a variant of bin packing where the items are
given from the beginning and variable-sized bins arrive online), \FFD
is better than \FFI~\cite{BF10}.
For seat reservation (the problem where a train with a certain number
of seats travels from station~$1$ to some station
$k \in \mathbb{Z}^+$, requests to travel from some station $i$
to a station $j>i$ arrive online,
and the aim is to maximize either the number of passengers or the total distance traveled), \firstfit and
\bestfit are better than \worstfit~\cite{BM08}
with regards to both objective functions.

\section{Approaches to Understanding Online Computation}
In this section, we discuss other means of analyzing and thereby
gaining insight into online computation. This includes
other performance measures and advice complexity.

\subsection{Other Performance Measures}
Other than competitive analysis, many alternative measures have been introduced
with the aim of getting a better or more refined picture of the (relative)
quality of online algorithms.

In chronological order, the main contributions are the following:
online/online ratio~\cite{GL90},
statistical adversary~\cite{R91},
loose competitive ratio~\cite{Y94},
max/max ratio~\cite{BB94},
access graphs (incorporating locality of reference)~\cite{BIRS95},
random-order ratio~\cite{K96},
accommodating ratio~\cite{BL99},
extra resource analysis~\cite{KP00},
diffuse adversary~\cite{KP00a},
accommodating function~\cite{BLN01},
smoothed analysis~\cite{ST04},
working set (incorporating locality of reference)~\cite{AFG05},
relative worst-order analysis~\cite{BF07,BFL07},
bijective and average analysis~\cite{ADLO07},
relative interval analysis~\cite{DLM09},
bijective ratio~\cite{ARS16},
and
online-bounded analysis~\cite{BEFLL16p,BEFLLtaj}.

We are not defining all of these measures here,
but we give some insight into the strengths and weaknesses of 
selected measures in the following. We start with a discussion
of work directly targeted at performance measure comparison.

\subsubsection{Comparisons of performance measures}
A systematic comparison of performance measures for online algorithms
was initiated in~\cite{BIL15}, comparing some measures which are
applicable to many types of problems. To make this feasible, a
particularly simple problem was chosen:
the $2$-server problem on a line with three points,
one
point farther away from the middle point than the other.

A well known algorithm, Double Coverage (\DC), is $2$-competitive and
best possible for this problem~\cite{CKPV91} according to competitive analysis.
A lazy version of this,
\LDC, is at least as good as \DC on every sequence and often better.
Investigating which measures can make this distinction,
\LDC was found to be better than \DC by bijective analysis and 
RWOA, but equivalent to \DC according to competitive analysis, the max/max
ratio, and the random-order ratio.
The first proof, for any problem, of an algorithm being best possible
under RWOA established this for \LDC .

\GREEDY performs unboundedly worse than \LDC on certain sequences, so ideally
a performance measure would not find \GREEDY to be superior to \LDC.
According to the max/max ratio and bijective analysis, \GREEDY
is the better algorithm, but not according to competitive analysis,
random-order analysis, or RWOA.

Further systematic comparisons of performance measures were made
in~\cite{BLM14} and~\cite{BLM15}, again comparing algorithms
on relatively simple problems.
The paper~\cite{BLM14} considered competitive analysis,
bijective analysis, average analysis, relative interval 
analysis, random-order analysis, and RWOA.
There were differences between the measures, but the most clear
conclusions were that bijective analysis found all algorithms
incomparable and average analysis preferred an
intuitively poorer algorithm.

Notable omissions from the investigations above are
extra resource analysis~\cite{KP00} and the accommodating function~\cite{BLN01},
both focusing on resources, which play a major r\^{o}le in most
online problems.
Both measures have been applied successfully to a range of problems,
giving additional insight; extra resource analysis (also referred to as resource augmentation) has been used
extensively. They can both be viewed as 
extensions of competitive analysis, explaining observed behavior of algorithms by expressing
ratios as functions of resource availability.

\subsection{Advice Complexity}
As a means of analyzing problems, as opposed to algorithms for those
problems, \emph{advice complexity} was proposed~\cite{DKP09,HKK10,BKKKM17}. 
The ``no knowledge
about the future'' property of online algorithms is relaxed, and it is
assumed that some bits of advice are available; such knowledge is 
available in many situations. One asks how many bits of advice are
necessary and sufficient to obtain a given competitive ratio,
or indeed optimality.
For a survey on advice complexity, see~\cite{BFKLM17}.

\section{Applicability}

Competitive analysis has been used for decades and sophisticated,
supplementary analysis techniques have been developed to make
proofs more manageable, or with the purpose of capturing more
fine-grained properties of algorithms.

We discuss two of the most prominent examples of these
supplementary techniques: \emph{list factoring} for analyzing
list accessing and \emph{access graphs} for modeling locality
of reference for paging.
Both techniques have been shown to work with RWOA.
As far as we know, list factoring has not been established as applicable
to any other
alternative to competitive analysis.
Access graphs have also been studied for relative interval analysis~\cite{BGL15}
with less convincing results.

\subsection{List Factoring for Analyzing List Accessing}

The idea behind \emph{list factoring} is to reduce the analysis to lists of two
elements, thereby making the analysis much more manageable.
The technique was first introduced by Bentley and
McGeoch~\cite{BM85} and later extended and improved~\cite{I91,T93,AvSW95,A98}.
In order to use this technique, one uses the \emph{partial cost model}, where
the cost of each request is one less than
in the standard (full) cost model (the access to the item itself is not counted). 
The list factoring technique is applicable for algorithms where, in treating any
request sequence $I$, one gets the same result by counting only the
costs of passing through $x$ or $y$ when searching for $y$ or $x$
(denoted $\ALG_{xy}(I)$),
as one would get if the original list contained only $x$ and $y$
and all requests different from those were deleted from the request
sequence, denoted $I_{xy}$.
If this is the case, that is $\ALG_{xy}(I) = \ALG(I_{xy})$ for all $I$, then
\ALG is said to have the \emph{pairwise property},
and it is not hard to prove that then
$\ALG(I) = \sum_{x \not= y} \ALG(I_{xy})$.
Thus, we can reduce the analysis of \ALG to an analysis of lists of length two.
The results obtained also apply in the full cost model if the algorithms
are \emph{cost independent}, meaning that their decisions are independent of the
costs of the operations.

Since the cost measure is different, some adaption is required to get
this to work for RWOA:

We now say that \ALG has the {\em worst-order projection property} if and
only if, for all sequences $I$, there exists a worst ordering
$\pi_{\ALG}(I)$ of $I$ with respect to \ALG, such that for all pairs
$\{a,b\} \subseteq L$ ($a \neq b$),
$\pi_{\ALG}(I)_{ab}$ is a worst ordering
of $I_{ab}$ with respect to \ALG on the initial list $L_{ab}$.

The results on \movetofront, \frequencycount, and
\transpose, reported on in Section~\ref{venividivici},
as well as results on \timestamp~\cite{A98}, 
were obtained~\cite{EKL13} using this tool.

\subsection{Access Graphs for Modeling Locality of Reference for Paging}
Locality of reference refers to the observed behavior of certain
sequences from real life, where requests seem to be far from uniformly
distributed, but rather exhibit some form of locality; for instance
with repetitions of pages appearing in close proximity~\cite{D68,D80}.
Performance measures that are worst-case over all possible sequences
will usually not reflect this,
so algorithms exploiting locality
of reference are not deemed better using the theoretical tools,
though they may be superior in practice.
This has further been underpinned by the following result~\cite{ADLO07} on
bijective analysis: 
For the class of demand paging algorithms (algorithms that never evict
a page unless necessary), for any two positive integers $m,n \in \mathbb{N}$,
all algorithms have the same number of input sequences of length~$n$
that result in exactly $m$ faults. 

One attempt at formalizing locality of reference, making it amenable
to theoretical analysis, was made in~\cite{BIRS95}, where
\emph{access graphs} were introduced.
An access graph is an undirected graph
with vertices representing pages and edges indicating that
the two pages being connected could be accessed immediately
after each other.
In the performance analysis of an algorithm, only sequences
respecting the graph are considered, i.e., any two distinct, consecutive
requests must be to the same page or to neighbors in the graph.

Under this restriction on inputs,
\cite{BIRS95,CN99} were able to show that, according to competitive analysis,
\LRU is strictly better than \FIFO on some access graphs
and never worse on any graph.
Thus, they were the first to obtain a separation,
consistent with empirical results.

Using RWOA, \cite{BGL12}~proved that
on the primary building blocks of access graphs, paths and cycles,
\LRU is strictly better than \FIFO.

\section{Open Problems and Future Work}
For problems where competitive analysis deems many algorithms best possible
or gives counter-intuitive results, comparing algorithms with RWOA can often
provide additional information. Such comparisons
can be surprisingly easy, since it is often possible to use
parts of previous results when applying RWOA.

Often the exploration for new algorithms for a given problem ends when
an algorithm is proven to have a best possible competitive ratio.
Using RWOA to
continue the search for better algorithms after competitive analysis fails to
provide satisfactory answers can lead to interesting discoveries.
As an example, the paging algorithm \RLRU was designed
in an effort to find an
algorithm that could outperform \LRU with respect to RWOA. 

Also for the paging problem, \RLRU and \LRUtwo are both known to be better
than \LRU according to RWOA. It was conjectured~\cite{BEKL10}
that \LRUtwo is comparable to \RLRU in \LRUtwo's favor.
This is still unresolved. It would be even more interesting to find a 
new algorithm better than both.
It might also be interesting to apply RWOA to an algorithm from the
class of \textsc{OnOpt} algorithms from~\cite{MN12}.

For bin packing, it would be interesting to
know whether \bestfit is better than \firstfit, according to RWOA.

\subsection*{Acknowledgment}
The authors would like to thank an anonymous referee for
many constructive suggestions.

\bibliographystyle{plain}
\bibliography{refs}

\end{document}